\documentclass[twocolumn,showpacs]{revtex4}

\usepackage{graphicx}%
\usepackage{dcolumn}
\usepackage{amsmath}

\begin{document}


\title{Resistivity of dilute 2D electrons in an undoped GaAs heterostructure}

\author{M. P. Lilly}
\email{mplilly@sandia.gov}
\author{J. L. Reno}
\author{J. A. Simmons}
\affiliation{Sandia National Laboratories, Albuquerque, NM 87185}%

\author{I. B. Spielman}
\author{J. P. Eisenstein}
\affiliation{California Institute of Technology, Pasadena, CA 91125}

\author{L. N. Pfeiffer}
\author{K. W. West}
\affiliation{Bell Laboratories, Lucent Technologies, Murray Hill, NJ 07974}

\author{E. H. Hwang}
\author{S. Das Sarma}
\affiliation{University of Maryland, College Park, MD 20742}

\date{\today}

\begin{abstract}
We report resistivity measurements from 0.03 K to 10 K in a dilute high
mobility 2D electron system. Using an undoped GaAs/AlGaAs heterojunction
in a gated field-effect transistor geometry, a wide range of densities,
$0.16 \times 10^{10} \mbox{cm}^{-2}$ to $7.5 \times 10^{10}
\mbox{cm}^{-2}$, are explored. For high densities, the results are
quantitatively shown to be due to scattering by acoustic phonons and
impurities. In an intermediate range of densities, a peak in the
resistivity is observed for temperatures below 1~K. This non-monotonic
resistivity can be understood by considering the known scattering
mechanisms of phonons, bulk and interface ionized impurities. Still
lower densities appear insulating to the lowest temperature measured.
\end{abstract}


\maketitle

The resistivity of 2D electrons at zero magnetic field has 
been used to probe scattering and quantum processes for many years.
Interest in the conducting behavior of 2D electrons and holes was
further heightened in 1994 when Kravchenko and coworkers discovered an
apparent metal-insulator transition of 2D electrons in Si
MOSFETs\cite{metal:si}. While 2D systems are expected to exhibit
insulating behavior (weak or strong localization) at sufficiently low
temperatures and/or densities, the metallic side of the metal-insulator
transition is more puzzling. 
A true metal conducts at $T=0$, however here we use the term
``metallic'' as referring to $d\rho/dT \ge 0$ at low
temperature, where $\rho$ is the resistivity.
Metallic behavior has been seen in a number of different
material systems, including electrons in Si MOSFETs\cite{metal:si},
holes in GaAs\cite{metal:pgaas1,metal:pgaas2,metal:pgaas3} and electrons
in GaAs\cite{metal:ngaas1,metal:ngaas2} provided the carrier density is
reasonably low (but not so low that the system is in the
insulating phase) and the mobility relatively high. While all of these
experimental systems exhibit metallic behavior in certain
regimes of density ($n$) and temperature, the quantitative
behavior varies widely. As the temperature is lowered, $\rho$ can 
decrease by a factor of 10 in
Si MOSFETs\cite{metal:si} compared to only a few percent for electrons in
GaAs\cite{metal:ngaas1}. The drop in resistivity is typically observed
when the dimensionless parameter $r_s$ (ratio of Coulomb to kinetic
energy, $\sim n^{-1/2}$) is much larger than 1, suggesting that
electron-electron interactions may be important. The issue
here is understanding the physical mechanisms underlying the
metallic behavior and elucidating the roles of disorder and
interaction. 

In this Letter, we present the temperature dependence of the resistivity
of a dilute 2D electron gas (2DEG) in GaAs. Electron-electron
interactions can be extremely important in this system at low temperatures
due to the low amount of disorder present in GaAs heterostructures. In
the sample discussed here, the density can be continuously tuned from
$0.16 \times 10^{10}$ cm$^{-2}$ ($r_s=13.7$) to $7.5 \times 10^{10}$
cm$^{-2}$ ($r_s=2.0$). With the ability to achieve very high $r_s$ in
low disorder samples, these resistivity measurements provide a test of
the importance of Coulomb interactions in the metallic regime. In
particular, the $r_s$ ($\stackrel{>}{_\sim} 10$) values of our 2DEGs are
very comparable to those in the best Si MOSFETs whereas the effective
disorder in our sample (as measured by $k_F \ell$, $k_F$
is the Fermi wavevector and $\ell$ the mean free path) is more than two
orders of magnitude lower. Earlier experimental
studies of the metallic behavior in 2D electrons in GaAs systems were
restricted to higher densities and lower mobilities, and
the drop in resistivity was much weaker than
the 20\% drop observed here.  Finally, the wide range of $n$ and $T$ reported here
allow several scattering regimes to be identified. 

We find that our measured temperature dependence of the resistivity can
be qualitatively well understood solely within the framework of Fermi
liquid theory. At high density, the resistivity is linear with
temperature, and the slope agrees both with previous
measurements\cite{phonons:exp1,phonons:exp2} and established theory for acoustic phonon
scattering\cite{phonons:sds}. For intermediate densities a low
temperature peak in the resistance is observed. The non-monotonic
resistivity agrees qualitatively with the predicted temperature
dependence of ionized impurity scattering (screening at low $T$ and the
transition from a degenerate Fermi gas to a classical system at high
$T$).

\begin{figure}
\includegraphics{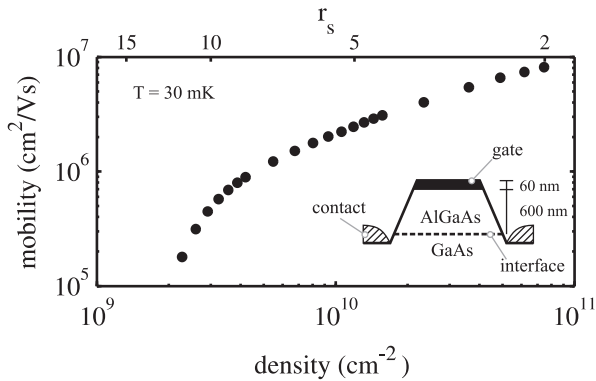}
\caption[fig1]{Plot of mobility as a function of electron density. 
The top axis indicates $r_s$. Inset is a schematic
cross section of the undoped heterojunction with self-aligned ohmic
contacts.}
\label{fig:mobility}
\end{figure}

The sample used in this study is a high mobility 2DEG confined at the
interface of a GaAs/Al$_{0.3}$Ga$_{0.7}$As heterojunction. The
interface (see Fig.~\ref{fig:mobility} inset)
is separated by undoped AlGaAs from a bulk doped GaAs cap
that serves as an integrated top gate.  A
$2 \times 2$~mm$^2$ square with 16 NiGeAu ohmic contacts is fabricated in a
field-effect transistor geometry\cite{kane}. The ohmic contacts are
self-aligned to the top gate, and carriers are drawn into the channel by
applying voltage between the gate and contacts.
Once present, the
density of the 2DEG is linearly proportional to the gate voltage. 
A calibration of the density is determined
using Shubnikov-de Haas oscillations for a number of gate voltages.
Although using undoped heterostructures requires complicated processing, 
an ultra-low density can be achieved without the penalty
of a substantial fixed disorder potential arising from remote delta-doping. 

The mobility, $\mu = 1/ne\rho$, is determined at
$T=30$~mK where we empirically observe that the
resistivity no longer changes with temperature. 
The mobility is extremely high
throughout the entire density range, as is shown in Fig.~\ref{fig:mobility}.  
At the
highest density for this sample the mobility is $8.2 \times
10^6$~cm$^2$/Vs ($k_F \ell = 2500$). In other devices that operate at
higher density, we have measured a mobility of $1.4 \times 10^7$
cm$^2$/Vs at a density of $2.5 \times 10^{11}$ cm$^{-2}$, a value 
to which the data in Fig.~\ref{fig:mobility} extrapolate.
Since the mobility continues to increase at high density, interface roughness
scattering is not relevant for data presented here\cite{ir_scattering}.

\begin{figure}
\includegraphics{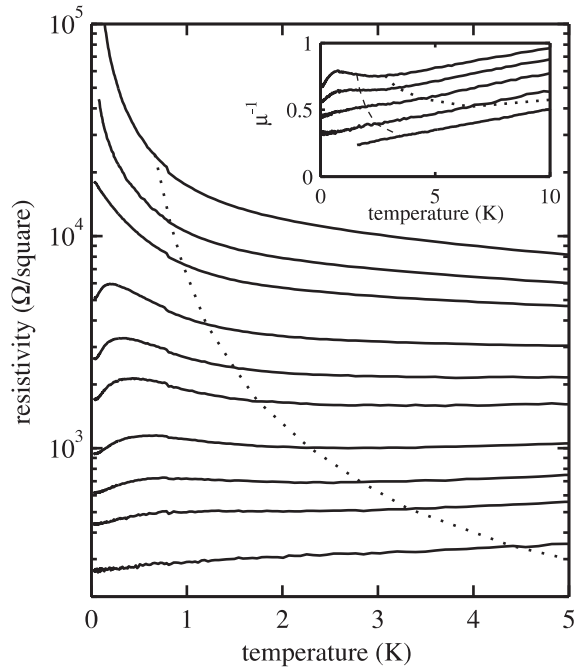}
\caption[fig2]{The main figure show the resistivity for densities (from
the top) 0.16, 0.20, 0.23, 0.29, 0.36, 0.42, 0.55, 0.68, 0.80 and 1.06
$\times 10^{10}$ cm$^2$.  The inset emphasizes the phonon component 
of the temperature dependence as straight parallel lines in inverse 
mobility (in units of $10^{-6}$ Vs/cm$^{2}$) 
for densities (from the top) 0.68, 0.80, 1.06, 1.57 
and 3.36 $\times 10^{10}$ cm$^{-2}$.  For clarity, the highest two 
densities in the inset are omitted from the main figure.  The dotted line
indicates $T_F$ and the dashed line in the inset indicates $T_{BG}$.
Note the log-scale in the main panel and the linear scale in the inset.}
\label{fig:rho}
\end{figure}

\begin{figure}
\includegraphics{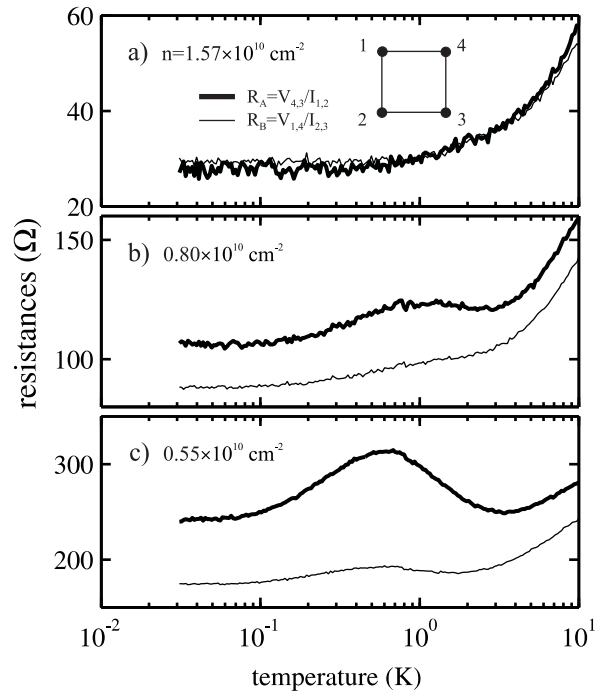}
\caption[fig3]{Four-wire resistances of the corner contacts
for densities shown.  The diagram labels the contacts.  
In defining $R_A$ and $R_B$, the 
subscripts indicate the current and voltage contacts.}  
\label{fig:r}
\end{figure}

The temperature dependence of the resistivity for a number of 
different densities is shown in
Fig.~\ref{fig:rho}. The locus of Fermi temperatures is
indicated with a dotted line. Three regimes can be identified. First, at
high density ($n>1.0 \times 10^{10}$ cm$^{-2}$) the resistivity is
constant at low temperature and then increases as $T$
increases. For intermediate densities ($0.29$ to $1.0 \times 10^{10}
\mbox{cm}^{-2}$) the resistivity is non-monotonic with temperature. 
As $T$ increases, $\rho$ is initially constant (see Fig.~\ref{fig:r}), 
then increases and decreases to form a local maximum below $T = 1$~K.
The third regime occurs at the lowest densities, with the
resistivity decreasing as the temperature increases in the
experimentally accessible temperature range.

The phonon contribution to the resistivity at high density is better
shown in the inset of Fig.~\ref{fig:rho}, where $\mu^{-1} = n e \rho$ is
plotted for $n=0.68$ to $3.36 \times 10^{10}$~cm$^{-2}$. The inverse
mobility for $T \stackrel{>}{_\sim} 2$~K is linear, and each density
has roughly the same
slope of $\sim 3.2 \times 10^{-8}$ Vs/Kcm$^2$). Acoustic
phonon scattering with both piezoelectric and deformation potential
coupling leads to a linear dependence of $\mu^{-1}$ above the
Bloch-Gr\"{u}neisen temperature ($kT_{BG} = 2k_F c \hbar$, $c$ is the
phonon speed of sound).  Previously reported
experimental\cite{phonons:exp1,phonons:exp2} and
theoretical\cite{phonons:sds} studies of higher density 2DEGs in
conventional GaAs heterostructures {\em quantitatively} agree with slope
found here. This demonstrates that the resistivity of the high density
2DEG is accurately captured using well known scattering mechanisms.
Note that the high temperature resistivity of the low density
data in the inset of Fig.~\ref{fig:rho} remains consistent with
phonon scattering even as the non-monotonic features begin to appear at low
temperatures. 

\begin{figure}
\includegraphics{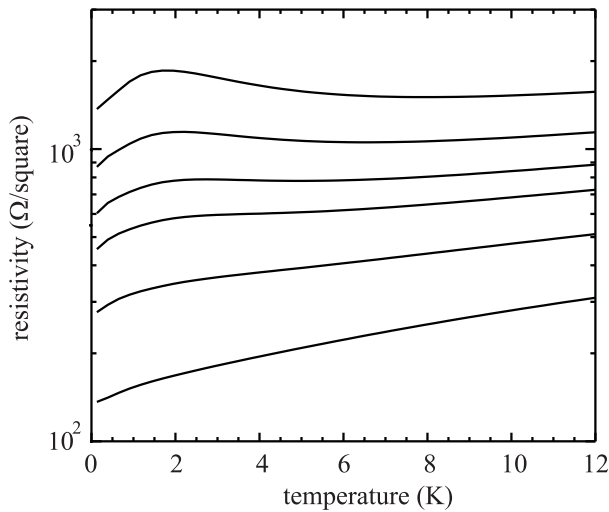}
\caption[fig4]{Result from scattering calculations are remarkably
similar to the data in Fig. 3. The density (from the top) is 0.42, 0.55
0.68,0.80, 1.06 and $1.57 \times 10^{10}$~cm$^{-2}$. Calculations
include acoustic phonons, bulk and interface impurities.}
\label{fig:theory}
\end{figure}

As the density is lowered below $n=1 \times 10^{10}$~cm$^{-2}$, the
resistivity becomes non-monotonic. A peak in $\rho$ appears for $T <
T_F$, and the position of the peak shifts to lower temperature for lower
densities. The resistivity at the maximum can be up to 23\% larger than
the low temperature ($T=30$~mK) value of the resistivity. The
non-monotonic peak occurs for $T<T_{BG}$, where phonon scattering is
strongly suppressed. In this regime of $n$ and $T$ the temperature
dependence of ionized impurity scattering must be considered. Screening
leads to $\rho(T) \sim T$ for $T \ll T_F$\cite{sds_ionized}, and the crossover from a
degenerate Fermi gas to a classical system leads to $\rho(T) \sim 1/T$
for $T>T_F$\cite{sds_ionized}. Competition between these effects will lead to a low $T$
resistivity peak. Detailed calculations including both ionized impurity
and acoustic phonon scattering are presented below
(Fig.~\ref{fig:theory}).

For the lowest density that can be attained in our samples $\rho$
decreases monotonically as $T$ increases. While it appears that the 2D
electron system is insulating for these densities, comparison to the
non-monotonic resistivities indicates 
a different interpretation is possible. Clearly, if the
data were only taken to 0.2~K, some of the non-monotonic curves at higher
density would also appear insulating.
Similarly, it is possible that the low $n$ insulating 
behavior could become non-monotonic if
the electron temperature were lowered further.  
From the results in Fig.~\ref{fig:rho}, it is clear that a separatrix 
cannot be identified in this system.
The functional form of the ``insulating'' temperature dependence for $n \le
0.23 \times 10^{10} \mbox{cm}^{-2}$ is more consistent with a power law
dependence than an activated or Mott variable range hopping behavior. 

In performing $\rho$ measurements at low $n$ great care must be taken to
insure that features do not arise as a result of inhomogeneities in the
current flow.
The temperature dependence of the resistances is shown in
Fig.~\ref{fig:r}.  Two resistances, $R_A$ and $R_B$, are
measured for current flowing, on average, in orthogonal directions. 
A constant current of 0.3~nA (for
the lowest density) to 20~nA (for the highest density) is used in a
standard lock-in amplifier measurement at a frequency of 7 Hz.
At these currents, transport measurements (e.g. quantum Hall minima) 
indicate that the electrons cool below $\sim$ 50~mK.
The resistivity is determined from the resistances using a procedure
prescribed by Van der Pauw\cite{vdp}. At high density ($n=1.57 \times
10^{10}$~cm$^{-2}$, Fig.~\ref{fig:r}a), the curves $R_A(T)$ and $R_B(T)$ 
almost lie on top of each other
and the temperature dependence is due to
phonon scattering. At lower density (Fig.~\ref{fig:r}b,c),
non-monotonic features appear in both $R_A$ and $R_B$, but the
resistances are no longer identical. Here we observe $R_A/R_B$ can be as large as 1.6.
Earlier designs of undoped samples
exhibited a severe ratio; $R_A/R_B$ approached 30. 
This effect originates in the
sample fabrication, with the higher resistance always related to the
location of the metallic gate contact. We speculate that the 
gate contact created a locally strained region above the square
2DEG causing inhomogeneous current flow. After redesigning the sample geometry to make the gate contact
far from the square measurement region, 
$R_A/R_B$ was reduced by more than a factor of 10. While this effect is
still visible in Fig.~\ref{fig:r}b and c, the remaining ratio 
has a very small ($<$ 10\%) impact on the
resistivity\cite{vdp}. Both $R_A$ and $R_B$ are non-monotonic, with 
resistance peaks located at nearly the same temperature.  
The qualitative nature of the non-monotonic resistivity is due to 
the underlying physics of ionized impurity scattering and not inhomogeneous
current flow.

To understand the low temperature ($T<1$~K) data, where the phonons
contribute little to the resistivity as the system enters the 
Bloch-Gr\"{u}neisen regime, we have carried out a microscopic transport
calculation using the Boltzmann theory, where we include effects of
charged impurity scattering on the electronic resistivity.  Our
calculation includes the following effects: (1) scattering by ionized
charged impurities in the bulk GaAs layer and at the GaAs/AlGaAs
interface; (2) temperature dependent screening, where the 
electron-charged impurity scattering is taken to be the temperature dependent
statically screened Coulomb interaction with the screening calculated
within the finite wavevector random phase approximation; (3) acoustic
phonon scattering including both deformation potential and piezoelectric
coupling; (4) the quasi-2D nature of the system through subband 
form-factors calculated in the Fang-Howard-Stern variational scheme.
This approach is valid for $k_F \ell > 1$.

We show our calculated results in Fig.~\ref{fig:theory} for several
experimental intermediate and high densities.
Reasonable values are chosen for bulk and interface impurities
to allow a comparison of the calculated results to the experimental
data shown in Fig.~\ref{fig:rho}.
Note that the overall
resistivity scale depends on the unknown impurity densities, but the
qualitative trends in $\rho(T,n)$ are real and arise from very basic
aspects of the underlying scattering mechanisms.  The rise in $\rho$
with increasing $T$ at low temperatures is a direct effect of the
thermal weakening of screening.  The non-monotonicity of $\rho(T)$ at
the intermediate temperatures arises from the competition among
temperature dependent screening, quantum-classical crossover and phonon
scattering.  The rise at higher $T$ is a phonon effect.

Quantitative understanding requires a more sophisticated theory which
includes higher order electron-electron interactions and disorder
induced localization corrections. Recently, electron-electron
interaction corrections have been considered by Zala and
coworkers\cite{zala}. They find that for a range of temperature,
$\hbar/\tau \ll T \ll T_F$ where $\tau$ is the transport relaxation
time, $\rho(T)$ is expected to be linear. From the slope, the Fermi
liquid parameter $F^0$ can be determined. In the data presented here
there is no significant range over which linear $\rho$ is observed for
$T \ll T_F$ (except for the phonon effect at higher $T$), and therefore
we do not attempt to determine $F_0$. 

In conclusion, we have made resistivity measurements on low disorder
gatable 2DEGs over a wide range of densities and temperatures. 
The resistivity data of high density 2DEGs agree
quantitatively with both experimental and theoretical accounts of
acoustic phonon scattering in GaAs. For the lowest densities, insulating
behavior is observed. At intermediate densities, the resistivity becomes
non-monotonic, with a peak appearing for $T<T_F$.  
At low temperature the competition between screening
($\sim T$) and a crossover from a degenerate Fermi system to a classical
2D gas ($\rho \sim T^{-1}$) leads to the formation of a peak.
Theoretical scattering
calculations including both ionized impurities and acoustic phonons in GaAs
show excellent qualitative agreement
with the data. This agreement between experiment and theory 
suggests that the underlying physics involved is
conventional Fermi liquid physics.

We acknowledge outstanding technical assistance from W. Baca and
R. Dunn.
This work has been supported by the Division of Materials Sciences and Engineering,
Office of Basic Energy Sciences, US Department of Energy.
Sandia is a multiprogram laboratory operated by Sandia
Corporation, a Lockheed Martin Company, for the United States Department
of Energy under contract DE-AC04-94AL85000.

\end{document}